\documentclass[conference,a4paper]{IEEEtran}
\usepackage[acronyms,nonumberlist,nopostdot,nomain,nogroupskip]{glossaries}
\usepackage{physics}

\usepackage{xcolor}
\usepackage{balance}

\usepackage[capitalise]{cleveref}

\ifCLASSINFOpdf
  \usepackage[pdftex]{graphicx}
\else
  \usepackage[dvips]{graphicx}
\fi
\usepackage{amsmath}
\ifCLASSOPTIONcompsoc
 \usepackage[caption=false,font=normalsize,labelfont=sf,textfont=sf]{subfig}
\else
 \usepackage[caption=false,font=footnotesize]{subfig}
\fi

\usepackage{enumitem}

\usepackage{tikz}

\newcommand{\sinr}{\ensuremath{\mathrm{SINR}}}
\newcommand{\sinrmean}{\ensuremath{\overline{\sinr}}}

\IEEEoverridecommandlockouts
\newcommand\copyrighttext{%
  \footnotesize \textcopyright 2020 IEEE. Personal use of this material is permitted.
  Permission from IEEE must be obtained for all other uses, in any current or future media, including reprinting/republishing this material for advertising or promotional purposes, creating new collective works, for resale or redistribution to servers or lists, or reuse of any copyrighted component of this work in other works.}
\newcommand\copyrightnotice{%
\begin{tikzpicture}[remember picture,overlay]
\node[anchor=south,yshift=10pt] at (current page.south) {\fbox{\parbox{\dimexpr\textwidth-\fboxsep-\fboxrule\relax}{\copyrighttext}}};
\end{tikzpicture}%
}

\newacronym{3gpp}{3GPP}{3rd Generation Partnership Project}
\newacronym{5g}{5G}{5th generation}
\newacronym{5gc}{5GC}{5G Core}
\newacronym{adc}{ADC}{Analog to Digital Converter}
\newacronym{aimd}{AIMD}{Additive Increase Multiplicative Decrease}
\newacronym{am}{AM}{Acknowledged Mode}
\newacronym{amc}{AMC}{Adaptive Modulation and Coding}
\newacronym{aqm}{AQM}{Active Queue Management}
\newacronym{awgn}{AGWN}{Additive White Gaussian Noise}
\newacronym{ard}{ARD}{Automatic Relevance Determination}
\newacronym{balia}{BALIA}{Balanced Link Adaptation}
\newacronym{bdp}{BDP}{Bandwidth-Delay Product}
\newacronym{bf}{BF}{Beamforming}
\newacronym{bs}{BS}{Base Station}
\newacronym{cc}{CC}{Congestion Control}
\newacronym{cdf}{CDF}{Cumulative Distribution Function}
\newacronym{cn}{CN}{Core Network}
\newacronym{cp}{CP}{Control Plane}
\newacronym{cqi}{CQI}{Channel Quality Information}
\newacronym{crs}{CRS}{Cell Reference Signal}
\newacronym{csirs}{CSI-RS}{Channel State Information - Reference Signal}
\newacronym{dc}{DC}{Dual Connectivity}
\newacronym{dce}{DCE}{Direct Code Execution}
\newacronym{dci}{DCI}{Downlink Control Information}
\newacronym{dl}{DL}{Deep Learning}
\newacronym{dmr}{DMR}{Deadline Miss Ratio}
\newacronym{dmrs}{DMRS}{DeModulation Reference Signal}
\newacronym{e2e}{E2E}{End-to-End}
\newacronym{ecn}{ECN}{Explicit Congestion Notification}
\newacronym{edf}{EDF}{Earliest Deadline First}
\newacronym{enb}{eNB}{evolved Node Base}
\newacronym{endc}{EN-DC}{E-UTRAN-\gls{nr} \gls{dc}}
\newacronym{epc}{EPC}{Evolved Packet Core}
\newacronym{es}{ES}{Edge Server}
\newacronym{fdd}{FDD}{Frequency Division Duplexing}
\newacronym{fdma}{FDMA}{Frequency Division Multiple Access}
\newacronym{fs}{FS}{Fast Switching}
\newacronym{ftp}{FTP}{File Transfer Protocol}
\newacronym{gnb}{gNB}{Next Generation Node Base}
\newacronym{gpr}{GPR}{Gaussian Process Regression}
\newacronym{harq}{HARQ}{Hybrid Automatic Repeat reQuest}
\newacronym{hpbw}{HPBW}{Half Power BeamWidth}
\newacronym{hetnet}{HetNet}{Heterogeneous Network}
\newacronym{hh}{HH}{Hard Handover}
\newacronym{hol}{HOL}{Head-of-Line}
\newacronym{hope}{HOpE}{High-dimensional OPtimization through Emulation}
\newacronym{hqf}{HQF}{Highest-quality-first}
\newacronym{ia}{IA}{Initial Access}
\newacronym{iab}{IAB}{Integrated Access and Backhaul}
\newacronym{imt}{IMT}{International Mobile Telecommunication}
\newacronym{inr}{INR}{Interference to Noise Ratio}
\newacronym{iot}{IoT}{Internet of Things}
\newacronym{kpi}{KPI}{Key Performance Indicator}
\newacronym{los}{LoS}{Line-of-Sight}
\newacronym{lte}{LTE}{Long Term Evolution}
\newacronym{m2m}{M2M}{Machine to Machine}
\newacronym{mac}{MAC}{Medium Access Control}
\newacronym{mc}{MC}{Multi-Connectivity}
\newacronym{mcs}{MCS}{Modulation and Coding Scheme}
\newacronym{mec}{MEC}{Mobile Edge Cloud}
\newacronym{mi}{MI}{Mutual Information}
\newacronym{mib}{MIB}{Master Information Block}
\newacronym{mimo}{MIMO}{Multiple-Input Multiple-Output}
\newacronym{ml}{ML}{Machine Learning}
\newacronym{mlr}{MLR}{Maximum-local-rate}
\newacronym{mlp}{MLP}{Multi-Layer Perceptron}
\newacronym[plural=\gls{mme}s,firstplural=Mobility Management Entities (MMEs)]{mme}{MME}{Mobility Management Entity}
\newacronym{mmwave}{mmWave}{millimeter wave}
\newacronym{mptcp}{MPTCP}{Multipath TCP}
\newacronym{mr}{MR}{Maximum Rate}
\newacronym{mrdc}{MR-DC}{Multi \gls{rat} \gls{dc}}
\newacronym{mss}{MSS}{Maximum Segment Size}
\newacronym{mtd}{MTD}{Machine-Type Device}
\newacronym{mtu}{MTU}{Maximum Transmission Unit}
\newacronym{nfv}{NFV}{Network Function Virtualization}
\newacronym{nlos}{NLoS}{Non-\gls{los}}
\newacronym{nr}{NR}{New Radio}
\newacronym{nrmse}{nRMSE}{normalized Root Mean Square Error}
\newacronym{nn}{NN}{Neural Network}
\newacronym{nyu}{NYU}{New York University}
\newacronym{nsa}{NSA}{Non Stand Alone}
\newacronym{o2i}{O2I}{Outdoor-to-Indoor}
\newacronym{ofdm}{OFDM}{Orthogonal Frequency Division Multiplexing}
\newacronym{pa}{PA}{Position-aware}
\newacronym{pbch}{PBCH}{Physical Broadcast Channel}
\newacronym{pdcch}{PDCCH}{Physical Downlink Control Channel}
\newacronym{pdcp}{PDCP}{Packet Data Convergence Protocol}
\newacronym{pdsch}{PDSCH}{Physical Downlink Shared Channel}
\newacronym{pdu}{PDU}{Packet Data Unit}
\newacronym{pf}{PF}{Proportional Fair}
\newacronym{pgw}{PGW}{Packet Gateway}
\newacronym{phy}{PHY}{Physical}
\newacronym{ppp}{PPP}{Poisson Point Process}
\newacronym{prb}{PRB}{Physical Resource Block}
\newacronym{pss}{PSS}{Primary Synchronization Signal}
\newacronym{pucch}{PUCCH}{Physical Uplink Control Channel}
\newacronym{pusch}{PUSCH}{Physical Uplink Shared Channel}
\newacronym{rach}{RACH}{Random Access Channel}
\newacronym{ran}{RAN}{Radio Access Network}
\newacronym[firstplural=Radio Access Technologies (RATs)]{rat}{RAT}{Radio Access Technology}
\newacronym{red}{RED}{Random Early Detection}
\newacronym{rf}{RF}{Radio Frequency}
\newacronym{rlc}{RLC}{Radio Link Control}
\newacronym{rlf}{RLF}{Radio Link Failure}
\newacronym{rmse}{RMSE}{Root Mean Square Error}
\newacronym{rr}{RR}{Round Robin}
\newacronym{rrc}{RRC}{Radio Resource Control}
\newacronym{rrm}{RRM}{Radio Resource Management}
\newacronym{rs}{RS}{Remote Server}
\newacronym{rsrp}{RSRP}{Reference Signal Received Power}
\newacronym{rsrq}{RSRQ}{Reference Signal Received Quality}
\newacronym{rss}{RSS}{Received Signal Strength}
\newacronym{rssi}{RSSI}{Received Signal Strength Indicator}
\newacronym{rtt}{RTT}{Round Trip Time}
\newacronym{rw}{RW}{Receive Window}
\newacronym{rx}{RX}{Receiver}
\newacronym{sa}{SA}{standalone}
\newacronym{sack}{SACK}{Selective Acknowledgment}
\newacronym{sap}{SAP}{Service Access Point}
\newacronym{sch}{SCH}{Secondary Cell Handover}
\newacronym{scm}{SCM}{Spatial Channel Model}
\newacronym{scoot}{SCOOT}{Split Cycle Offset Optimization Technique}
\newacronym{sdma}{SDMA}{Spatial Division Multiple Access}
\newacronym{si}{SI}{Study Item}
\newacronym{sib}{SIB}{Secondary Information Block}
\newacronym{sinr}{SINR}{Signal to Interference plus Noise Ratio}
\newacronym{sm}{SM}{Saturation Mode}
\newacronym{snr}{SNR}{Signal to Noise Ratio}
\newacronym{son}{SON}{Self-Organizing Network}
\newacronym{srs}{SRS}{Sounding Reference Signal}
\newacronym{ss}{SS}{Synchronization Signal}
\newacronym{sss}{SSS}{Secondary Synchronization Signal}
\newacronym{svm}{SVM}{Support Vector Machine}
\newacronym{svr}{SVR}{Support Vector Regressor}
\newacronym{tb}{TB}{Transport Block}
\newacronym{tcp}{TCP}{Transmission Control Protocol}
\newacronym{tdd}{TDD}{Time Division Duplexing}
\newacronym{tdma}{TDMA}{Time Division Multiple Access}
\newacronym{tfl}{TfL}{Transport for London}
\newacronym{tm}{TM}{Transparent Mode}
\newacronym{trp}{TRP}{Transmitter Receiver Pair}
\newacronym{tti}{TTI}{Transmission Time Interval}
\newacronym{ttt}{TTT}{Time-to-Trigger}
\newacronym{tx}{TX}{Transmitter}
\newacronym{ue}{UE}{User Equipment}
\newacronym{ul}{UL}{Uplink}
\newacronym{ula}{ULA}{Uniform Linear Array}
\newacronym{um}{UM}{Unacknowledged Mode}
\newacronym{umi}{UMi}{Urban Micro-cell}
\newacronym{uml}{UML}{Unified Modeling Language}
\newacronym{upa}{UPA}{Uniform Planar Array}
\newacronym{utc}{UTC}{Urban Traffic Control}
\newacronym{ut}{UT}{User Terminal}
\newacronym{vm}{VM}{Virtual Machine}
\newacronym{vr}{VR}{Virtual Reality}
\newacronym{wbf}{WBF}{Wired Bias Function}
\newacronym{wf}{WF}{Wired-first}
\makeglossaries

\begin{document}
%
\title{Machine Learning-aided Design of\\Thinned Antenna Arrays\\for Optimized Network Level Performance\vspace{-0.45cm}}

\author{{  \parbox{\linewidth}{ \centering
\textbf{Mattia Lecci},
\textbf{Paolo Testolina},
\textbf{Mattia Rebato},
\textbf{Alberto Testolin},
\textbf{Michele Zorzi}}}\\
 Department of Information Engineering, University of Padova, Italy \\
\{\texttt{leccimat, testolina, rebatoma, testolin, zorzi}\} \texttt{@dei.unipd.it}\\
\vspace{-.45cm}
}



\maketitle
\copyrightnotice

\begin{abstract}

With the advent of \gls{mmwave} communications, the combination of a detailed 5G network simulator with an accurate antenna radiation model is required to analyze the realistic performance of complex cellular scenarios.
However, due to the complexity of both electromagnetic and network models, the design and optimization of antenna arrays is generally infeasible due to the required computational resources and simulation time.
In this paper, we propose a Machine Learning framework that enables a simulation-based optimization of the antenna design.
We show how learning methods are able to emulate a complex simulator with a modest dataset obtained from it, enabling a global numerical optimization over a vast multi-dimensional parameter space in a reasonable amount of time.
Overall, our results show that the proposed methodology can be successfully applied to the optimization of thinned antenna arrays.

\end{abstract}


\begin{tikzpicture}[remember picture,overlay]
\node[anchor=north,yshift=-10pt] at (current page.north) {\parbox{\dimexpr\textwidth-\fboxsep-\fboxrule\relax}{\centering \footnotesize This paper has been presented at EuCAP 2020. \textcopyright 2020 IEEE.\\
  Please cite it as: M. Lecci, P. Testolina, M. Rebato, A. Testolin, and M. Zorzi, ``Machine Learning-aided Design of Thinned Antenna Arrays for Optimized Network Level Performance,'' 14th European Conference on Antennas and Propagation (EuCAP 2020), Copenhagen, Mar. 2020}};
\end{tikzpicture}%

\section{Introduction}
\label{sec:introduction}
Massive \glspl{upa} operating in the \gls{mmwave} frequency range will be adopted in the 5th generation of mobile networks (5G) as the key enablers to meet the challenging requirements of the new standard.
Large antenna arrays can compensate for the propagation and penetration losses at such high frequencies thanks to beamforming techniques, synthesizing 3D beams that can focus the transmitted power towards specific users~\cite{rangan14}, increasing the antenna gain and thus increasing the received power.
Furthermore, beamforming can help exploit the unique propagation characteristics of the \gls{mmwave} channel, such as spatial sparsity, to reduce the interference among users.

These ambitious goals require a novel approach to antenna design and optimization. 
Antenna arrays can no longer be designed and optimized without considering the network topology: in addition to the common antenna design goals, such as decreasing the side-lobe level or maximizing the directivity, more global, network-oriented requirements need to be taken into account.
Such requirements dramatically increase the complexity of the optimization problem, as it moves from the bare electromagnetic to the network domain.
As both antenna prototyping and network deployment tests are prohibitively expensive for both academia and most industry, electromagnetic and network simulators are often employed.
In~\cite{rebato18}, the accurate modeling of antennas in network simulators was proved to be decisive, further confirming that design and optimization need to carried out jointly.

\begin{figure}
\centering
\includegraphics[width=1\columnwidth]{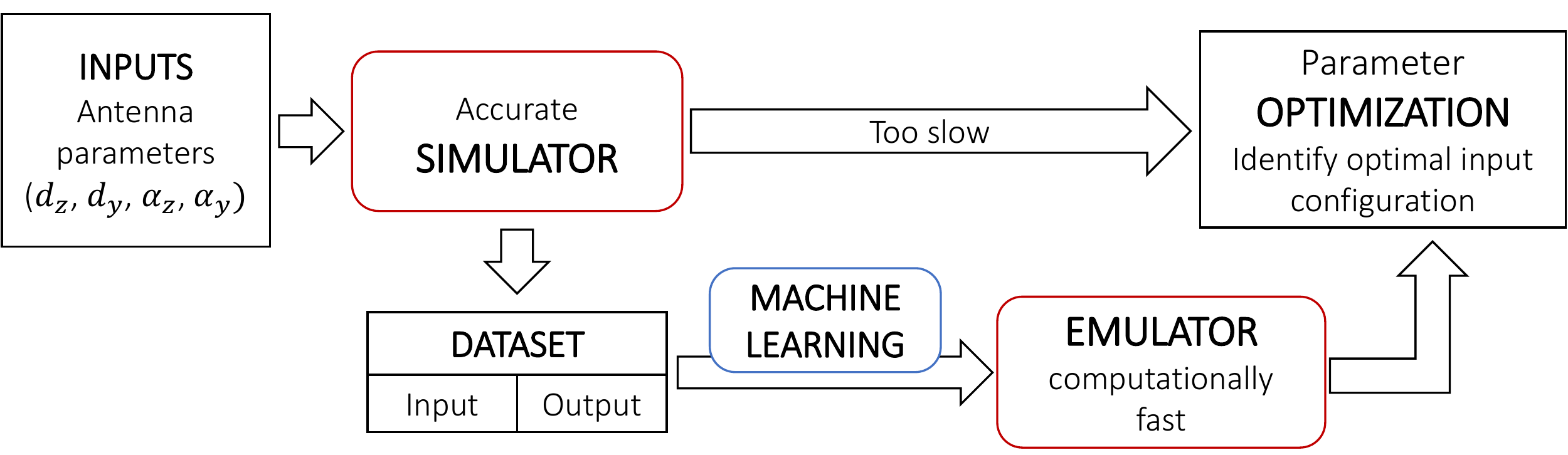}
\setlength\abovecaptionskip{-.3cm}
\caption{Workflow of the proposed framework.
The diagram highlights how the parameter optimization is achieved using an ML-based emulator.
\vspace{-3ex}}
\label{fig:workflow}
\end{figure}

Heuristic simulation-based optimization is generally not feasible, as such detailed simulators are both time and computationally expensive.
Indeed, the large number of iterations needed by optimization algorithms prevents the use of simulations requiring hours (or even days) of running time.
For this reason, in this paper, we propose and evaluate a \gls{ml} framework that can mimic a given simulator and allows us to achieve any network optimization objective in a reasonable amount of time.
The general framework is represented in \cref{fig:workflow} and has been recently described in detail elsewhere~\cite{Testolina2019}.
The diagram shows how the parameter optimization can be achieved through the \gls{ml}-based emulator, which only requires a single training phase done using a dataset of simulated data.

In this work, differently from~\cite{Testolina2019}, the focus is on the optimization of thinned arrays.
This manuscript is organized as follows: in \cref{sec:simulator_description} the new design challenge is described; in \cref{sec:framework} new learning techniques are tested for the emulation task, to tackle the increased complexity of the problem; finally, in \cref{sec:results} we present the results of the optimization, while \cref{sec:conclusions} concludes this work and describes some remaining open challenges.

\subsection{Related Works}
\label{sec:related_works}

Recently, \gls{ml} techniques have started to be applied as a tool to solve many kinds of problems.
Also in the communication field, there exist many works adopting learning-oriented approaches to address complex transmission issues.  
Particular attention has been gathered by the new database proposed in~\cite{Alkhateeb2019}, as it lays the premises for a common research ground.

One common application of \gls{ml} is parameter estimation, where great results were achieved even where the most sophisticated classical techniques failed.
An example can be found in~\cite{Arnold2019}, where the authors try to estimate the downlink channel starting from samples of the uplink channel.
While well-known signal processing techniques (e.g., the Wiener filter) were not able to perform a good estimate, the \gls{ml} approach proposed by the authors yielded very good results.

\gls{ml} has been successfully applied also at the network layer.
Innovative ideas and proposals have challenged even the most resilient classical paradigms such as the ISO/OSI architecture~\cite{cognetworks}.
These new approaches started showing their potential in the increasingly heterogeneous network scenarios, e.g., when facing the high data load and quality of experience required for video streaming~\cite{testolin}.

Moreover, the authors in~\cite{sun18} use Deep \glspl{nn} to optimize the allocation algorithm in a wireless resource management problem.
The proposed concept is similar to the one described in our work, as a learning tool is used to approximate a complex input-output function.
However, the authors also include the optimization step into the learning process and use many more training samples to accommodate the needs of their deep architecture.

Considering that at high frequencies, such as in the \gls{mmwave} bands, strong attenuations are present, quantifying the actual antenna gain obtained due to the radiation pattern is fundamental to precisely evaluate any \gls{mmwave} system. 
For this reason, we have used previous works~\cite{rebato18,Testolina2019} as the main references for both the network and the antenna characterizations.

Finally, regarding the specific problem of thinned arrays, several works exist on their optimization at the antenna level.
A reference for the general theory and results can be found in~\cite{surveySen}. 
On the optimization side,~\cite{thinningLi} and~\cite{thinningSingh} apply genetic algorithms to the activation mask of the array to further lower the side-lobe level.
Nevertheless, as mentioned above, none of these works consider network metrics in the antenna design.
Some works, such as~\cite{Haupt2010}, explicitly employ thinning for interference reduction, but they do not rely on simulated network statistics that allow achieving ad hoc optimizations, tailored to the network characteristics.
This is due to the high complexity of the network simulators, which would be added to the already high computational load of the electromagnetic simulation.
Here is where our framework can prove itself useful, providing an agile emulator that can be used to speed up complex optimization tasks to reasonable execution time.

\section{Simulator Description}
\label{sec:simulator_description}
As in~\cite{Testolina2019}, a \gls{3gpp} compliant simulator was used to extract network-level metrics, such as \gls{sinr} statistics, based on a Monte Carlo approach.
Specific environment parameters follow the \gls{3gpp} standards~\cite{3gpp.38.901,3gpp.38.913} based on the \gls{umi} scenario with no \gls{o2i} losses.

The goal of this study is to understand whether irregular thinning is a desirable property in an array.
In \cref{sub:antenna_array_generation} we describe the adopted irregular thinning approach, while in \cref{sub:scenario_parameters} we list the parameters to be optimized.

\subsection{Antenna Array Generation}
\label{sub:antenna_array_generation}

\begin{figure}[tbp]
  \vspace{-0.2cm}
  \centering
  \includegraphics[width=0.45\textwidth]{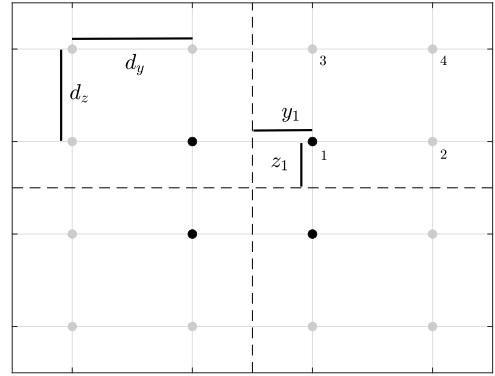}
  \setlength\abovecaptionskip{-.3cm}
  \caption{Example of a generated array.
  Dashed lines separate the four quadrants, while black and gray dots represent respectively the activated antennas and the array lattice.
  The top-left quadrant is generated and then mirrored to the other three.
  \vspace{-2ex}}
  \label{fig:irregularly_thinned_array}
\end{figure}
\begin{figure*}[tbp]
  \centering
  \includegraphics[width=0.9\textwidth]{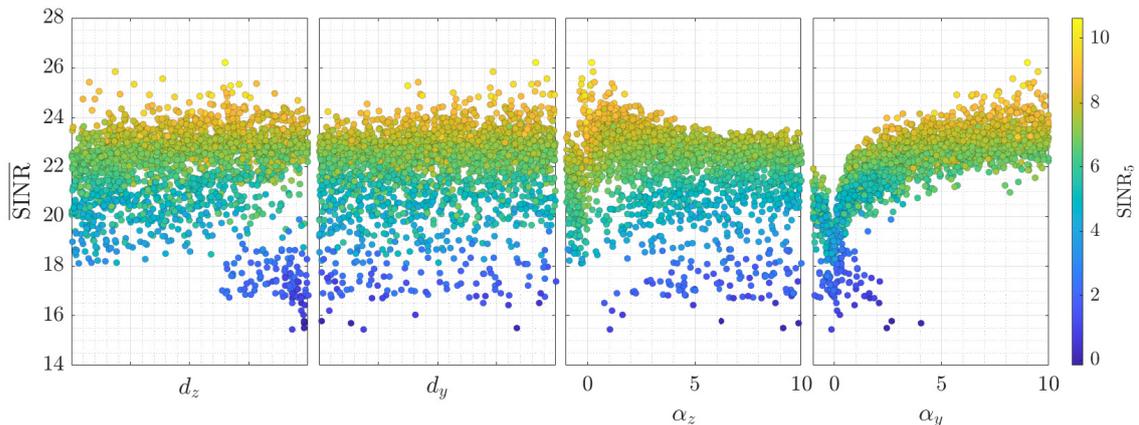}
  \setlength\abovecaptionskip{-.3cm}
  \caption{Correlation plot of the four input parameters vs the output metric ($y$-axis) and the bound metric (color).
  From here it is already possible to see the importance of the antenna shape parameters for the system performance, giving a hint on the optimal parameters found by a good emulator.
  \vspace{-3ex}}
  \label{fig:pairplot}
\end{figure*}

To simplify both the implementation and the optimization, thinning is defined by means of an \emph{activation mask} over a regular lattice of dummy antennas.
Namely, a large antenna array lattice is created but only some of the antennas are turned on (see \cref{fig:irregularly_thinned_array}).
Thus, all the antenna elements have approximately the same element pattern and thinned arrays are more easily parameterized.

The activation mask is randomly produced at each iteration of the Monte Carlo simulation as follows.
First, the lattice is split into four quadrants.
Then, starting from the center of the lattice, a \emph{probability profile} $f(\Delta_y, \Delta_z) = f_y(\Delta_y) f_z(\Delta_z)$ is defined, where $\Delta_y$ and $\Delta_z$ are the distances of the antenna elements from the center of the lattice in the horizontal and vertical dimensions, respectively. 
Considering a single quadrant, each element $i = 1, \ldots, N_{quadrant}$ in position $(y_i,z_i)$ is assigned a value $v_i = u_i \cdot f(y_i,z_i)$, where $\{u_i\}$ are i.i.d. uniform random variables defined in the interval $[0,1]$.
Finally, the elements with the largest values $v_i$ are chosen and the sample quadrant is mirrored over the other three, to force a realistic symmetry.
In this paper, a probability profile following an exponential decay $f_y (\Delta_y)= e^{-\alpha_y \Delta_y}$ (and analogously for $f_z$) is chosen.

\subsection{Scenario Parameters}
\label{sub:scenario_parameters}

Specific values and ranges were chosen based on our previous experiences~\cite{Testolina2019}, to optimize the positioning of $64$ antenna elements over a given lattice.
Results shown in \cref{sec:results} are based on a fixed lattice with $100 \times 99$ antenna elements spaced apart by $d_y,\, d_z$.
Regarding the generation of the activation mask, the probability profile is parameterized by $\alpha_y,\, \alpha_z \in [-1,10]$.
Values are chosen to allow a very wide range of possibilities, including extremely sparse ones.
Please note that increasing values of $\alpha$ tend to push active antennas together towards the center, while negative values tend to push them towards the outer edges of the lattice.

\section{Framework Definition}
\label{sec:framework}
\glsreset{ml}
The objective of the proposed framework is to greatly restrict the search for an optimal input configuration to a small subspace or, if possible, to find the globally optimal configuration, speeding up simulation-based optimization in the presence of slow simulators.
Specifically, in this work, the framework is applied to the network simulator described in \cref{sec:simulator_description}.
Due to its complexity, using it for a brute-force optimization would be extremely inefficient.
Instead, \gls{ml} algorithms are trained to \emph{learn} the simulator's input-output relationship.
Once the \emph{emulator} is trained, the network statistics can be computed for any input configuration almost instantaneously.
In this way, unseen configurations can be emulated extremely quickly during the optimization phase, thus dramatically reducing the time required for the optimization. 

\subsection{Data Analysis}
\label{sub:data_analysis}

A preliminary data analysis is customary in \gls{ml} problems, as it helps analyze correlations in the given dataset, hinting towards the selection (or exclusion) of some learning algorithms.
It can also serve both as a sanity check for the optimization results and to discover unexpected data distributions and anomalies.

One of the most basic yet useful tools in high dimensional feature spaces is the correlation plot, i.e., a matrix of scatter plots showing the correlation among variables, as reported in \cref{fig:pairplot}.
As expected, some trends are easily recognizable.
Nevertheless, visual inspection alone cannot be exhaustive due to the high dimensionality of the input space, nor can the presence of some minima/maxima in the dataset guarantee the global optimality of such points.
For research purposes, we generated a total of $N=1,000$ random configuration, each obtained with $10,000$ Monte Carlo iterations.

\subsection{Learning Methods}
\label{sub:ml}

The problem we are facing is a numerical regression on synthetic, noisy data.
We recall that the input feature space is a four-dimensional hyperspace where: 
\begin{itemize}
  \item $d_z, d_y$ are the vertical and horizontal spacing, expressed as fractions of $\lambda$;
  \item $\alpha_y,\, \alpha_z \in [-1,10]$ are the probability profile parameters.
\end{itemize}
The algorithm will predict all the outputs that are required by the optimization process, including the ones used for the constraints. 
For instance, if one wants to optimize the inputs with respect to variable A (e.g., mean \gls{sinr}, \sinrmean) while bounding variable B (e.g., minimum coverage requirement on the $5^{th}$ percentile of the \gls{sinr}, $\sinr_5$), the emulator must be able to predict both A and B.

Several algorithms were tested, but only a selected subset will be hereby described.
The performance of the different techniques is evaluated using 5-fold cross-validation, and, as we are interested in keeping the training set as small as possible, the comparison is made for different training set sizes. 
Thus, it is possible to know the accuracy of the emulation, based on the number of available samples.
\begin{itemize}
    \item \emph{Linear regression} is the most basic class of regression algorithms. 
    Despite its simplicity, many versions and adaptations have been created, able to solve non-trivial problems.
    It is often considered as a baseline for more powerful algorithms.
    Adding a ridge regularization to the linear regression helps avoid overfitting the training data by imposing a penalty on the size of the weights.
    \item \emph{Random forests} are ensembles of decision trees, that approximate stepwise the target function;
    \item \emph{\glspl{svr}} are derived from the \gls{svm} classification algorithm.
    Among all the typical kernels, the Gaussian one performed best and is used here.
    \item \emph{\glspl{ard}} directly derive from Bayesian Ridge Regression, but includes a sparsity assumption in the priors which stabilizes the weights;
    \item \emph{\gls{mlp}} is a well-known architecture that should be able to approximate any function.
    Nevertheless, \glspl{mlp} generally require (i) long and computationally-demanding hyperparameter tuning and (ii) large datasets.
\end{itemize}
The performance is evaluated using the \gls{nrmse} metric, as in \cite{Testolina2019}.
Considering a scalar output $y$, the prediction or emulation error is then computed as the difference between the prediction of the emulator $\hat{y}$ and the corresponding simulator output $y$, normalized with respect to the latter, namely
\begin{equation}
\label{nrmse_def}
\mathrm{nRMSE} = \sqrt{\frac{1}{N} \sum_{i=1}^{N}
    \qty( \frac{y_i-\hat{y}_i}
    {y_i} )^2 }
\end{equation}
Results are reported in \cref{fig:incremental}.

\section{Optimization Results}
\label{sec:results}

\begin{figure}[tbp]
  \centering
  \includegraphics[width=0.45\textwidth]{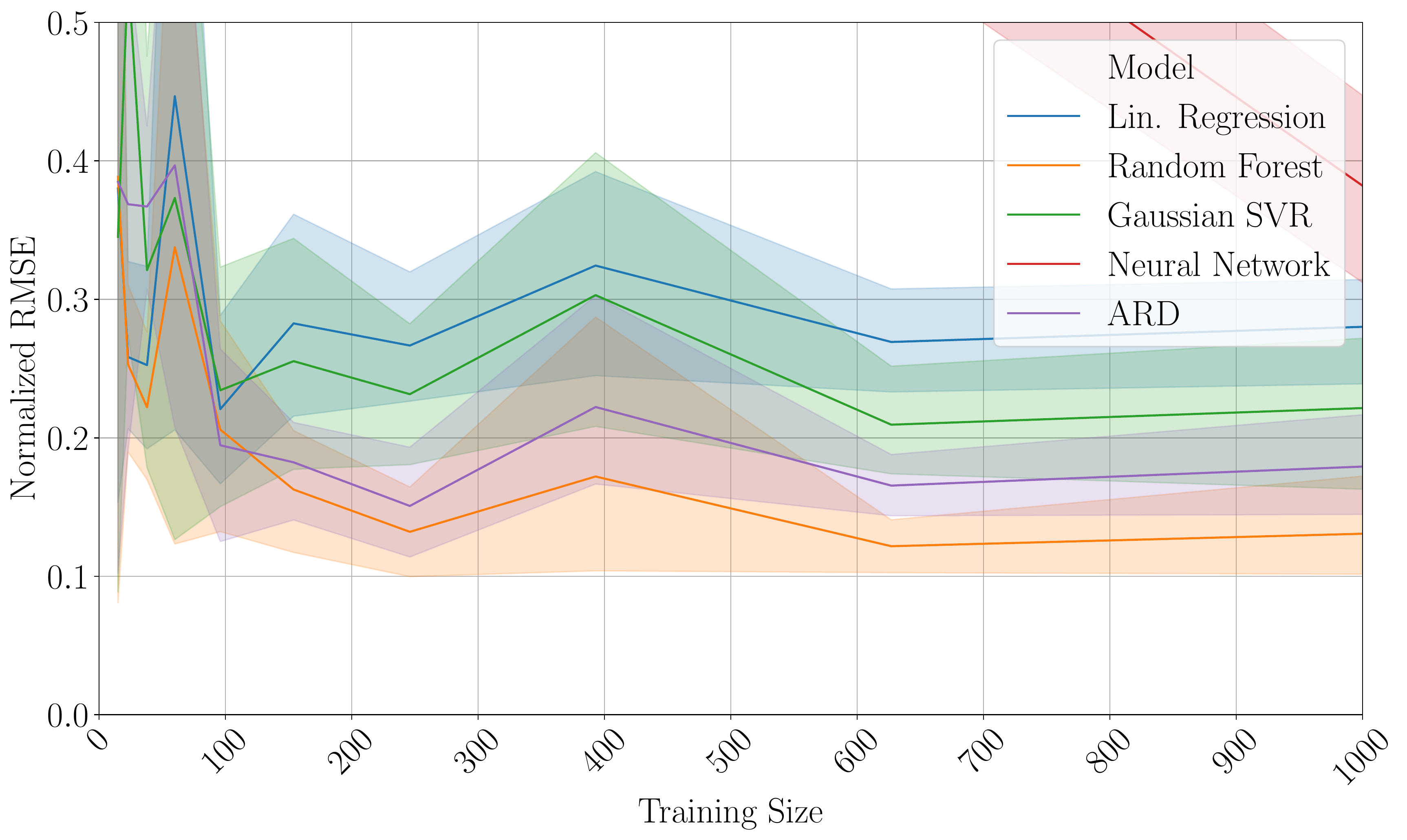}
  \setlength\abovecaptionskip{-.1cm}
  \setlength\belowcaptionskip{-.3cm}
  \caption{Plot showing cross-validation scores on the \gls{nrmse} metric with increasing training size.
  \vspace{-2ex}}
  \label{fig:incremental}
\end{figure}

The algorithms described is \cref{sub:ml} were evaluated for increasing dataset sizes.
During the creation of the dataset, monitoring the learning performance as the number of available samples increases can help find a plateau of the learning process, allowing to stop the simulations when the required precision is achieved.
In fact, the prediction performance is fundamentally limited by the noise in the given dataset, mainly caused by the limited number of Monte Carlo iterations.
As the prediction residuals are symmetrically distributed around zero, this should not affect the generalization performance of the model.
Based on the comparison of the described algorithms in \cref{fig:incremental}, we decided to use \emph{Random Forests} for our emulator as they give the best results even with as few as $500$ data points.

The objective function chosen for this problem optimizes the average performance of the given network (mean \sinr{}) while imposing a minimum coverage level, corresponding to a lower bound to the $5^{th}$ percentile of the \sinr{} as follows
\begin{equation}
\begin{aligned}
  &\text{maximize} &&\sinrmean\\
  &\text{subject to} && \sinr_5 > 6 \text{ dB}
\end{aligned}
\end{equation}

Given the description from \cref{sub:antenna_array_generation}, it should be noted that the outcome of this optimization problem does not yield the best possible antenna for the given scenario, but rather a family of antennas following a probability distribution obtained for the optimal parameters $\alpha_y^*$, $\alpha_z^*$, together with the optimal lattice spacing $d_z^*$, $d_y^*$.

For comparison, we consider as the baseline antenna an $8\times 8$ \gls{upa} with $d_z=d_y=0.5\lambda$ spacing.
Also, we compare the results with the optimal antenna previously found in~\cite{Testolina2019}, given by a vertical linear array of $64$ elements, with $d_z=0.796\lambda$.

\begin{figure}[tbp]
  \vspace{-0.2cm}
  \centering
  \includegraphics[width=0.45\textwidth]{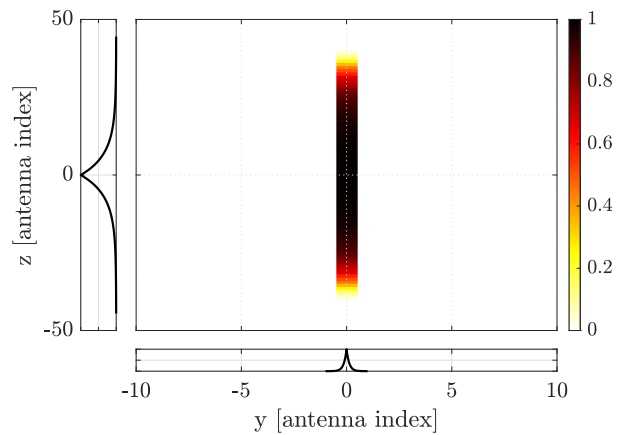}
  \setlength\abovecaptionskip{-.2cm}
  \caption{Visual representation of the activation probability of any given element from the lattice using the optimal parameters $\alpha_y^*$, $\alpha_z^*$.
  Elements outside the central column are never activated, indicating that a vertically-shaped antenna is optimal.
\vspace{-0.5cm}}
  \label{fig:activationMaskStats}
\end{figure}

The \gls{ml}-based optimization framework suggests as the optimal parameters $\alpha_z^*=9.02$, $\alpha_y^*=0.20$, $d_z^*=0.761$, $d_y^*=0.866$.
To better understand what these parameters suggest, \cref{fig:activationMaskStats} shows the probability that any given antenna in the lattice is active, together with the corresponding probability profiles.
It can be easily noted that the activation probability indicates that vertical antennas tend to perform better than any other configuration, similarly to what was found in our previous work where the $64\times 1$ configuration was identified as optimal.

\begin{figure}[tbp]
  \centering
  \includegraphics[width=0.45\textwidth]{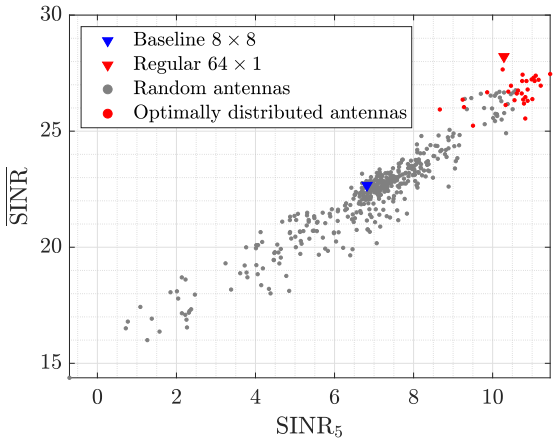}
  \setlength\abovecaptionskip{-.1cm}
  \caption{Performance comparison between different classes of antennas.
    Gray dots are obtained choosing $300$ antennas generated from non-optimized input configurations from the input space described in \cref{sub:scenario_parameters}.
    Red dots, instead, show $30$ antennas, all generated with the optimal input configuration.
    Blue and red triangles represent the baseline and the optimal antenna found in~\cite{Testolina2019}, respectively.
    \vspace{-0.3cm}}
  \label{fig:cdfComparison}
\end{figure}

As these results do not identify a specific antenna, but rather a family of antennas, in \cref{fig:cdfComparison} we show a comparison between (i) two specific antennas used as references, (ii) antennas generated using non-optimized (i.e., randomly selected) input parameters, and (iii) antennas generated using the optimal ones.
Note that, while the antennas from the optimal family do not perform equally, they always achieve significantly better performance with respect to the baseline and to the other configurations and closely approach the optimal antenna found in~\cite{Testolina2019}, often improving over the $\sinr_5$ although not over the \sinrmean.
Though the input parameters optimized by the framework do not directly identify a specific antenna configuration, they allow to drastically reduce the search space to a much narrower area, that can be further explored using more traditional, time-consuming techniques.

\begin{figure}[tbp]
  \centering
  \vspace{-0.1cm}
  \includegraphics[width=0.45\textwidth]{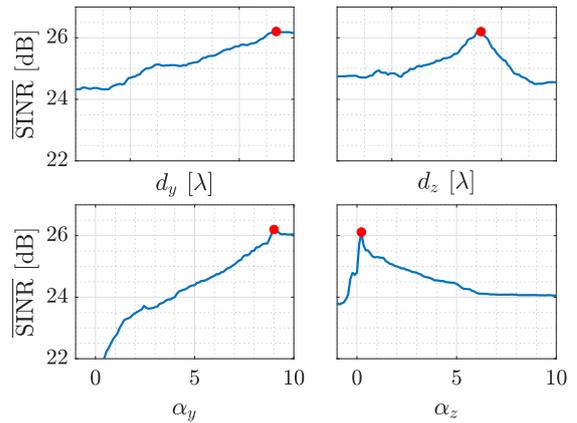}
  \setlength\abovecaptionskip{-.3cm}
  \caption{Visualization of the input configuration around the optimal value.
    Slices of the input configuration are taken to show how the different parameters affect the performance.
    It is clear that the antenna shape parameters $\alpha_y$, $\alpha_z$ play a more important role than the lattice-related parameters $d_y$, $d_z$.
    \vspace{-0.5cm}}
  \label{fig:slicePlots}
\end{figure}

Finally, we can study the sensitivity of our optimal point with respect to the four input parameters in \cref{fig:slicePlots}.
As expected, $\alpha_y^*$ is chosen to be large, forcing the antenna to be vertical.
Instead, while a large value of $\alpha_z^*$ would push the elements towards the center to make it less sparse, it would also tend to make the antenna more rhomboidal.
The optimization was thus able to find the largest possible value for the vertical sparsity that still allowed all antennas to be strictly vertical.
Given the preference for a vertical antenna, the horizontal spacing $d_y^*$ is the parameter with the least effects on the network performance.
The vertical spacing $d_z^*$ is instead similar to the one previously found.

\section{Conclusions}
\label{sec:conclusions}

In this paper, we showed that an \gls{ml}-based optimization framework can be successfully used to optimize antenna design in a very efficient way. 
Thanks to the antenna parameterization chosen in this study, our framework was able to explore much more complex configurations than regularly spaced planar arrays.
Returning an optimized family of antennas rather than a specific configuration successfully reduces the search space of possible configurations, making it possible to further refine it with more precise simulations.
Finally, in a 3GPP-compliant Urban Micro-cell scenario with static users, the optimizer suggests that vertical linear arrays are the optimal configuration, supporting the results of our previous work.

\section*{Acknowledgment}

This project was funded by Huawei.
We would like to thank Jonathan Gambini, Laura Resteghini, and Christian Mazzucco (\textit{Huawei Research, Milan}) for their support and collaboration.
Mattia Lecci's and Paolo Testolina's activities were partially supported by \emph{Fondazione Cassa di Risparmio Padova e Rovigo} under the grant ``Dottorati di Ricerca'' 2018 and 2019, respectively.

\bibliographystyle{IEEEtran}
\bibliography{bibl}

\balance

\end{document}